\documentclass[pdflatex,sn-mathphys-num]{sn-jnl}

\usepackage[T1]{fontenc}
\usepackage[english]{babel}
\usepackage{multirow}
\usepackage{array}
\usepackage[section]{placeins}
\usepackage{graphicx}
\usepackage{dcolumn}
\usepackage{bm}
\usepackage{amssymb}
\usepackage{ulem}

\usepackage{mathrsfs,amsbsy,amstext}
\usepackage{amsmath}
\usepackage{setspace}
\usepackage{esint}
\usepackage{xcolor,colortbl}
\usepackage{booktabs}
\usepackage{tikz}
\usepackage{braket}
\usepackage{enumitem}
\usepackage{listings}
\usepackage{xcolor}

\usepackage{physics}
\usepackage{comment}
\usepackage{soul}

\usetikzlibrary{%
  positioning,
  arrows.meta
}

\begin{document}

\title[Quantum feedback algorithms for DNA assembly using FALQON variants]{Quantum feedback algorithms for DNA assembly using FALQON variants}

\author*[1]{\fnm{Pedro} \sur{M. Prado}}\email{pm.prado@unesp.br}

\author[2]{\fnm{Lucas} \sur{A. M. Rattighieri}}

\author[1]{\fnm{Rafael} \sur{Simões do Carmo}}

\author[1]{\fnm{Giovanni} \sur{S. Franco}}

\author[1]{\fnm{Guilherme} \sur{E. L. Pexe}}

\author[3]{\fnm{Alexandre} \sur{Drinko}}

\author[3]{\fnm{Erick} \sur{G. Dorlass}}

\author[3]{\fnm{Tatiana} \sur{F. de Almeida}}

\author*[1,3]{\fnm{Felipe} \sur{F. Fanchini}}\email{felipe.fanchini@unesp.br}

\affil[1]{\orgname{S\~ao Paulo State University (UNESP), School of Sciences}, \orgaddress{\postcode{17033-360}, \state{Bauru-SP}, \country{Brazil}}}
\affil[2]{\orgname{Instituto de Física Gleb Wataghin, Universidade de Campinas}, \orgaddress{\postcode{13083-970}, \state{Campinas-SP}, \country{Brazil}}}
\affil[3]{\orgname{Hospital Israelita Albert Einstein}, \orgaddress{\state{São Paulo-SP}, \country{Brazil}}}

\abstract{Reconstructing DNA sequences without a reference, known as \textit{de novo} assembly, is a complex computational task involving the alignment of overlapping fragments. To address this problem, a usual strategy is to map the assembly to a Quadratic Unconstrained Binary Optimization (QUBO) formulation, which can be solved by different quantum algorithms. In this work, we focus on standard Feedback-based Algorithm for Quantum Optimization (FALQON) and two previously proposed feedback-based variants, which update control parameters without an external classical parameter-optimization routine. We analyze DNA sequence fragments from SARS-CoV-2 and human mitochondrial DNA using standard FALQON, second-order FALQON (SO-FALQON), and time-rescaled FALQON (TR-FALQON). Numerical results show that both variants can improve convergence and success probability while requiring fewer feedback layers than standard FALQON in selected configurations. These results support the use of enhanced feedback dynamics as a proof-of-concept strategy for reduced DNA-assembly instances, while practical scalability and hardware performance remain open questions.}


\maketitle

\section{Introduction}\label{intro}

Genomic diversity is driven by complex events such as recombination, duplication, and transposition. While these mechanisms are essential for evolution, they render the genome computationally challenging to reconstruct. In biotechnology, the \textit{de novo} reconstruction of genomes from high-throughput sequencing \textit{reads} has become a critical step for the characterization of organisms without prior reference, as well as for metagenomic studies and clinical diagnostics \cite{10.1093/bib/bbw096}. These reads, typically hundreds of bases long, present partial overlaps that must be optimally aligned, a task complicated by genomic repetitions, sequencing errors, and the large volume of data. These challenges have motivated classical graph-based and heuristic assembly approaches, including de Bruijn graph methods \cite{Zerbino2008} and assemblers such as Velvet, SPAdes, and ABySS \cite{Bankevich2012,Simpson2009}, although large and repetitive genomes remain computationally challenging.

High-fidelity genome assemblies enable fundamental applications in medicine and biology such as the identification of genetic variants associated with hereditary diseases, monitoring of infectious outbreaks through genomic surveillance, studies of molecular evolution, and the development of personalized therapies and regenerative medicine \cite{Venter2001,Shendure2017}. In addition, fast and accurate assemblies reduce research costs and enable real-time analysis in critical clinical situations.

The formulation of the assembly problem as a \textit{Travelling Salesman Problem} (TSP) on a graph whose edges represent negative overlap costs naturally leads to the QUBO (Quadratic Unconstrained Binary Optimization) expression, an approach that benefits from advances in quantum optimization and hybrid simulations \cite{Lucas2014,Glover2018,Jain2021,carmo2025warm,carmo2025warmpce}. QAOA (Quantum Approximate Optimization Algorithm) has previously been applied to a QUBO formulation of DNA assembly \cite{QuASeR2024}.

 Quantum approaches to DNA reconstruction have also been investigated using quantum annealing \cite{NaleczCharkiewicz2022,NatureQA2021} and graph-based pangenome formulations related to de Bruijn graph representations \cite{cudby2025pangenome}.

Concurrently with the development of the present work, a closely related study proposed a quantum-assisted graph-optimization pipeline for \textit{de novo} assembly, formulating Hamiltonian and Eulerian path problems on the assembly graph as a higher-order binary optimization (HOBO) model solved with a gate-based VQE, and introducing a bitstring-recovery mechanism to improve exploration of the solution space \cite{pamidimukkala2026accelerating}. This independent progress reinforces the timeliness of quantum-assisted formulations for assembly, while highlighting complementary directions: whereas that approach couples quantum optimization with classical pre-/post-processing, here we focus on feedback-driven quantum dynamics via FALQON and its variants, without an external classical parameter-optimization loop.

Within gate-based quantum optimization, variational quantum algorithms (VQAs) constitute a widely explored framework for solving optimization problems on near-term quantum devices \cite{Cerezo2021}. QAOA is a representative example of this approach \cite{Farhi2014}. At circuit depth $p$, QAOA alternates evolutions generated by the problem and driver Hamiltonians, controlled by two sets of variational parameters, $\boldsymbol{\gamma}=(\gamma_1,\ldots,\gamma_p)$ and $\boldsymbol{\beta}=(\beta_1,\ldots,\beta_p)$. These parameters are determined through an external classical optimization routine that repeatedly evaluates the quantum objective function and updates the parameter vector. Consequently, increasing the circuit depth also increases the number of variational parameters and the dimensionality of the associated classical optimization problem. Feedback-based quantum optimization provides a different strategy in which control parameters are instead updated directly from expectation values measured from the quantum system according to a prescribed feedback law. FALQON belongs to this class of methods and employs quantum Lyapunov control to progressively drive the system toward low-energy states of the problem Hamiltonian \cite{Magann_2022}.

An important practical aspect of these approaches is that expectation values are estimated from a finite number of circuit measurements. As a consequence, the estimated objective functions and observables are affected by statistical fluctuations associated with measurement shot noise. In variational quantum optimization, this uncertainty can increase the number of circuit repetitions required to obtain sufficiently accurate estimates and may affect the performance of the classical optimization routine \cite{Scriva2024,Barligea2025,Kungurtsev2024}. In FALQON, finite-sampling effects arise in a different way. Although the algorithm does not employ an external classical parameter-optimization routine, the expectation values used in the feedback law must also be estimated from a finite number of measurements. Therefore, shot noise can directly affect the determination of the control parameters and, consequently, the evolution of the algorithm.

The notion of quantum speedup in optimization must therefore be interpreted carefully. Quantum optimization includes exact, approximate, and heuristic approaches, and the evidence required for a quantum advantage depends on the computational setting and on the classical methods used as reference \cite{Abbas2024}. In heuristic NISQ algorithms, an improvement in convergence or a reduction in circuit depth relative to another quantum method does not, by itself, demonstrate a quantum speedup. A practical advantage requires comparisons that account for both solution quality and computational resources, including time to solution, circuit execution, and sampling cost \cite{Lykov2023}. Accordingly, the present work does not claim a quantum speedup over classical optimization methods; instead, it investigates whether feedback-based modifications can improve convergence and reduce circuit depth relative to standard FALQON for the considered DNA-assembly instances.

In the present work, we investigate the performance of FALQON applied to the DNA assembly problem and compare it with two previously proposed variants. Alongside standard FALQON, we consider two previously proposed variants: the \textit{Time-Rescaled FALQON} (TR-FALQON), which applies temporal rescaling to the circuit operators to accelerate convergence at shallow depths~\cite{Rattighieri2025}; and the \textit{Second-Order FALQON} (SO-FALQON), which modifies the feedback law by including a second-order approximation in the time interval, allowing larger steps and a significant reduction in circuit depth~\cite{Arai2025}. Both variants were developed to reduce the number of feedback layers required for convergence, a feature that is relevant to depth-constrained quantum implementations.

This paper is organized as follows: in Section~\ref{sec:dna} we discuss in detail the DNA assembly problem and its rigorous mapping to the QUBO/Ising formalism; in Section~\ref{sec:falqon} we present the motivation, Lyapunov control, and functioning of the FALQON algorithm; in Section~\ref{sec:tr-falqon} we describe the TR-FALQON variant, with emphasis on temporal rescaling techniques and their benefits for shallow circuits; Section~\ref{sec:so-falqon} is dedicated to the formulation of SO-FALQON and the discussion of the second-order approximation; in Section~\ref{sec:implementacao} we present the computational implementation of the methods and experiments; in Section~\ref{sec:resultados} we report and analyze the results obtained, comparing energy convergence, control parameters, and success probabilities of the different approaches; and finally, in Section~\ref{sec:conclusao} we discuss conclusions and perspectives for the work.

\section{DNA Decoding Problem via de novo Assembly}
\label{sec:dna}

De novo genome assembly aims to reconstruct a complete genomic sequence without the aid of a pre-existing reference, starting from small sequencing fragments generated by high-throughput sequencing platforms such as Illumina. Each read has a typical length that varies depending on the technology and may contain insertion, deletion, and substitution errors \cite{Shendure2017}. The coverage depth (the average number of times each genomic position is read) must be sufficient to guarantee redundancy but approximately linearly increases the data volume. Additionally, repetitive regions and structural variants create ambiguities that hinder exact assembly, requiring robust methods to distinguish real biological events from sequencing artifacts \cite{Zerbino2008,Bankevich2012}.

\subsection{Overlap--Layout--Consensus Model}
\label{subsec:olc}

In the overlap--layout--consensus paradigm, each sequencing fragment (read) \(r_i\) is represented by a vertex in a directed graph
\[
G = (V,E), \quad V = \{r_1,r_2,\dots,r_N\}.
\]

This model, introduced in Sanger fragment assemblies and refined for long reads of third-generation sequencing \cite{Pevzner2001,Myers2005}, remains the basis of modern assemblers such as Canu and Flye \cite{Berlin2015,Kolmogorov2019}.

For each ordered pair \((r_i,r_j)\), the maximum overlap length for which the suffix of \(r_i\) matches the prefix of \(r_j\) is defined as \cite{Li2016Minimap},

\begin{align}
\mathrm{ov}(i,j)
=
\max_{k \le \min(|r_i|,|r_j|)}
\bigl\{
\text{suffix of }r_i\text{ of length }k
=
\text{prefix of }r_j\text{ of length }k
\bigr\}.
\end{align}

The adjacency matrix elements of this directed graph become

\begin{align}
w_{i,j}=-\mathrm{ov}(i,j),
\end{align}
where more negative values correspond to larger overlaps and, therefore, greater assembly affinity \cite{QuASeR2024,Zerbino2008}. Note that, in general,

\[
\mathrm{ov}(i,j)\neq\mathrm{ov}(j,i),
\]
because the suffix--prefix comparison is directional; sequencing errors can further modify these asymmetric overlap scores \cite{Myers2005}.

\medskip

For real sequencing data, the overlap between two reads should account for substitution, insertion, and deletion errors \cite{Li2010,Marinier2015Pollux}. A substitution produces a local mismatch between aligned bases, whereas insertions and deletions shift the relative positions of the subsequent bases. As a consequence, a correct overlap may not be identified through direct position-by-position comparison.

Before performing a complete approximate alignment, candidate read pairs can be selected using a $q$-gram filter. A $q$-gram is a substring of length $q$, and two sequences within a bounded edit distance must share a minimum number of such substrings. Therefore, pairs that do not satisfy this condition can be discarded before the more computationally expensive alignment step \cite{Jokinen1991,UKKONEN1992191}. Similar $q$-gram filtering strategies are commonly used to reduce the search space in sequence alignment and overlap detection.

For the remaining candidate pairs, the suffix of $r_i$ and the prefix of $r_j$ can be compared using approximate alignment. This procedure searches for the alignment with the best score while allowing mismatches and gaps. Substitutions appear as different bases at the same aligned position, whereas insertions and deletions are represented by gaps. The alignment score can assign positive contributions to matches and different penalties to mismatches and gaps. Thus, a small number of sequencing errors reduces the quality of the overlap without necessarily eliminating it.

The weight $w_{i,j}$ can then be defined from the resulting alignment score, considering both the length of the aligned region and the errors identified. Ambiguous nucleotides can also be treated at this stage. IUPAC symbols represent sets of possible bases, such as \texttt{R}, which represents \texttt{A} or \texttt{G}, and \texttt{N}, which represents an undetermined base \cite{CornishBowden1985Nomenclature}. These symbols can contribute to the alignment score according to their compatibility with the corresponding base.

The presence of sequencing errors can modify the relative overlap values and produce different pairs of reads with similar scores, increasing the ambiguity of the overlap graph. As a consequence, different read orderings may become associated with similar total overlap scores, making it more difficult to distinguish the optimal assembly from alternative configurations. A quantitative assessment under realistic sequencing-error models is left for future work. In the instances considered in this work, however, the overlap values are calculated from error-free reads using exact suffix--prefix matches. These overlap values define the edge weights of the graph and therefore determine the cost associated with each possible ordering of the reads.

The search for a minimum-cost Hamiltonian cycle in $G$ then aims to find an order of visiting all vertices that maximizes the sum of consecutive overlaps. This cycle arises because each vertex must be used exactly once, forming a loop that returns to the starting point or extending into a linear sequence, depending on the assembly strategy. In optimization terms, a Hamiltonian cycle corresponds to the solution that simultaneously satisfies the constraint of single use of each read and organizes the overlaps in a globally optimal manner \cite{Simpson2009,Pevzner2001}.

\noindent\textbf{Simplified example:}

As an illustrative example, we use the following four reads, constructed as cyclic rotations of length 8 to guarantee nonzero overlaps for all pairs:

\[
\begin{aligned}
r_{0} &= \texttt{ATCGATCG},\\
r_{1} &= \texttt{CGATCGAT},\\
r_{2} &= \texttt{TCGATCGA},\\
r_{3} &= \texttt{GATCGATC}.
\end{aligned}
\]

The overlap graph constructed from these four reads is presented in Fig.~\ref{fig:overlap_graph}.

\begin{figure}[h]
  \centering
  \begin{tikzpicture}[>=stealth,scale=1.5,every node/.style={font=\small}]
    \node[circle,draw,inner sep=2pt] (r0) at (0,1) {$r_0$};
    \node[circle,draw,inner sep=2pt] (r1) at (1,0) {$r_1$};
    \node[circle,draw,inner sep=2pt] (r2) at (0,-1) {$r_2$};
    \node[circle,draw,inner sep=2pt] (r3) at (-1,0) {$r_3$};

    \draw[->] (r0) to[bend left=20] (r1);
    \draw[->] (r1) to[bend left=20] (r0);
    \draw[->] (r0) to[bend left=10] (r2);
    \draw[->] (r2) to[bend left=10] (r0);
    \draw[->] (r3) to[bend left=20] (r0);
    \draw[->] (r0) to[bend left=20] (r3);
    \draw[->] (r1) to[bend left=30] (r2);
    \draw[->] (r2) to[bend left=10] (r1);
    \draw[->] (r3) to[bend left=20] (r1);
    \draw[->] (r1) to[bend left=20] (r3);
    \draw[->] (r2) to[bend left=20] (r3);
    \draw[->] (r3) to[bend left=20] (r2);

    \node[anchor=west] at (2.0,0) {
        $W=\left(\begin{matrix}
        0 & -6 & -7 & -5\\
        -6 & 0 & -5 & -7\\
        -5 & -7 & 0 & -6\\
        -7 & -5 & -6 & 0
        \end{matrix}\right)$
    };
  \end{tikzpicture}

  \caption{Directed complete graph constructed from four reads and its corresponding adjacency matrix. Each matrix element $(i,j)$ represents the directed overlap weight from read $r_i$ to read $r_j$.}
  \label{fig:overlap_graph}
\end{figure}

After identifying the minimum-cost Hamiltonian cycle, the \textit{consensus} phase is applied, where each pair of overlapping reads is aligned according to the vertex sequence of the cycle, discrepancies are corrected by majority voting, and conflicts in repetitive regions are resolved based on coverage and quality \cite{Myers2005}. The result is a reconstructed genomic sequence, free of reference bias and with global maximization of overlaps.

\subsection{Problem Formulation as QUBO}
\label{sec:qubo}

The mapping of the de novo assembly problem to QUBO form allows its solution either by quantum annealing or by gate-based quantum algorithms such as variational or feedback-based methods \cite{Lucas2014,Glover2018,Choi2008Embedding}. This representation captures the combinatorial nature of the fragment reordering problem, formulating it as a quadratic energy minimization over binary variables subject to structural constraints.

Let $N$ be the number of reads after preprocessing (redundancy removal, quality filtering, and chimera detection), and let a binary variable $x_{i,p}\in\{0,1\}$ indicate whether read $r_i$ occupies position $p$ in the final assembly:

\[
x_{i,p}
=
\begin{cases}
1, & \text{if } r_i \text{ is at position } p,\\
0, & \text{otherwise},
\end{cases}
\qquad
i,p=0,\dots,N-1.
\]

This one-hot representation introduces $N^2$ binary variables in the general formulation, and therefore a quadratic encoding overhead with respect to the number of reads. In the instances considered in this work, one read is fixed at the first position to remove the cyclic symmetry of the assembly, reducing the number of required variables to $(N-1)^2$. Although this reduction is useful for small instances, the asymptotic scaling remains quadratic. Consequently, the dimension of the corresponding quantum state space grows as $2^{(N-1)^2}$, which rapidly limits the size of instances that can be represented directly. This encoding cost is particularly relevant in genomic applications, where biologically realistic datasets may contain very large numbers of reads, and can offset potential computational advantages provided by the optimization algorithm itself \cite{Maurizio2025}.

The total QUBO objective function is represented as

\begin{align}
\min_{\mathbf{x}\in\{0,1\}^{N^2}}
\mathbf{x}^{\top}Q\mathbf{x},
&\nonumber\\
Q
&=
A Q_{\text{pos}}
+
B Q_{\text{read}}
+
C Q_{\text{overlap}},
\label{eq:qubo}
\end{align}

where $\mathbf{x}$ is a column vector of binary variables $x_{i,p}$. Each component of Eq.~\eqref{eq:qubo} is defined as follows:

\begin{itemize}

\item \textbf{Position constraint:} Each position must contain exactly one read. The term penalizes any ambiguity or omission of allocation:

\[
Q_{\text{pos}}
=
\sum_{p=0}^{N-1}
\left(
1-\sum_{i=0}^{N-1}x_{i,p}
\right)^2.
\]

This term expands to generate quadratic and bilinear contributions that enforce uniqueness per position.

\item \textbf{Uniqueness constraint:} The same read cannot occupy more than one position:

\[
Q_{\text{read}}
=
\sum_{i=0}^{N-1}
\left(
1-\sum_{p=0}^{N-1}x_{i,p}
\right)^2.
\]

This term also expands to penalize multiple occurrences of the same read.

\item \textbf{Overlap cost:} For each consecutive pair of positions $(p,p+1)$, the cost is reduced by overlaps between reads,

\begin{align*}
Q_{\text{overlap}}
&=
\sum_{p=0}^{N-2}
\underset{i\neq j}{\sum_{i=0}^{N-1}\sum_{j=0}^{N-1}}
w_{ij}\,x_{i,p}\,x_{j,p+1}.
\end{align*}

\end{itemize}

The penalty coefficients $A$ and $B$ must satisfy

\[
A,B>\max_{i,j}|w_{ij}|,
\]
so that no violation of constraints can be compensated by an overlap gain. However, values of $A,B$ that are too large widen the spectral range of the Hamiltonian and hinder the solution process.

For quantum devices based on Pauli $\sigma_z$ operators, such as those used in quantum annealing or in gate-based quantum algorithms, it is necessary to convert the binary variables $x_{i,p}\in\{0,1\}$ into spin variables $z_a\in\{-1,+1\}$. This conversion is carried out by the linear transformation

\[
x_{i,p}
\to
\frac{1-z_a}{2}.
\]

Substituting this relation into the quadratic objective function, we obtain

\begin{align*}
\mathbf{x}^{\top}Q\mathbf{x}
&=
\left(\frac{1-z}{2}\right)^{\top}
Q
\left(\frac{1-z}{2}\right),\\
&=
\frac14 z^{\top}Jz
-
\frac12 h^{\top}z
+
\text{constant},
\end{align*}
where the coefficients are defined as

\[
J_{ab}=4Q_{ab},
\qquad
h_a=2\sum_b Q_{ab}.
\]

The added constant does not affect the optimal solution but is relevant for recovering the total energy. The final Hamiltonian in the Ising formalism takes the form

\[
H_{\text{Ising}}
=
\sum_{a<b}J_{ab}Z_aZ_b
+
\sum_a h_a Z_a,
\]
where $Z_a$ represents the Pauli $\sigma_z$ operator acting on qubit $a$. This form is directly compatible with implementation in quantum hardware, allowing optimization algorithms to exploit the topology and physical limitations of the underlying architecture.

Therefore, the QUBO/Ising mapping used in this work should be understood as a direct encoding suitable for investigating the behavior of feedback-based algorithms on reduced DNA-assembly instances, rather than as an immediately scalable representation for complete genomic datasets. The resource cost associated with the encoding must be considered together with the performance of the optimization algorithm when assessing the practical scalability of the approach.

One possibility for reducing this cost is to exploit the sparsity of the overlap graph. In OLC \textit{pipelines}, low-confidence overlaps, contained \textit{reads}, and redundant or transitive edges can be removed before the \textit{layout} stage. As a result, the graph used for optimization can be considerably sparser than the complete graph associated with the standard TSP formulation. For acyclic components, an edge-based QUBO formulation can assign one binary variable to each retained overlap, reducing the number of variables from $O(N^2)$ to $O(|E|)$ \cite{NatureQA2021}. When the graph degree remains bounded, this number can approach $O(N)$. This reduction, however, depends on the structure of the graph. Repetitive regions, ambiguous overlaps, and circular genomes can generate cycles and may require additional constraints or graph-decomposition techniques.

Another strategy is to modify the encoding of the ordering problem itself. More compact TSP formulations can reduce the number of required qubits to $O(N\log N)$ when higher-order binary interactions are allowed \cite{glos2022space}. In this case, the reduction in the number of qubits comes at the cost of higher-order interactions, so the formulation is no longer strictly QUBO and may increase the complexity of the Hamiltonian and the corresponding quantum circuit.

A third possibility is to use multibasis encodings. In this approach, binary variables are no longer represented exclusively through computational basis occupations associated with $Z$ operators. Instead, they are encoded through expectation values of Pauli operators and Pauli strings involving $X$, $Y$, and $Z$, as well as their permutations. Examples include $X^{\otimes k}\otimes \mathbb{I}^{\otimes(n-k)}$, $Y^{\otimes k}\otimes \mathbb{I}^{\otimes(n-k)}$, and $Z^{\otimes k}\otimes \mathbb{I}^{\otimes(n-k)}$, where $k$ denotes the number of correlated Pauli operators and $n$ the number of qubits. By associating optimization variables with these expectation values rather than only with computational basis states, a larger number of binary variables can be represented using a smaller number of qubits, leading to a polynomial compression in the required quantum resources \cite{sciorilli2025towards}. The treatment of constraints in this representation is less direct, although recent proposals have explored strategies for incorporating them \cite{carmo2025warmpce,padin2026pauli}.

These strategies suggest that a scalable approach to quantum-assisted genome assembly will likely require a combination of classical preprocessing, graph simplification, and more resource-efficient quantum encodings. Rather than representing all \textit{reads} simultaneously in a single dense QUBO, quantum optimization could be applied selectively to graph components in which assembly ambiguity remains after classical processing. The investigation of these alternatives is left for future work.

\section{FALQON Algorithm}
\label{sec:falqon}

\subsection{Motivation}

In the context of DNA sequence assembly, the problem can be mapped into a QUBO model and subsequently transformed into an Ising Hamiltonian, as discussed in Sec.~\ref{sec:qubo}. Traditionally, variational algorithms such as the QAOA \cite{Farhi2014} apply parameterized circuits whose angles must be externally optimized by classical routines
which can introduce a significant optimization cost, particularly in NISQ (Noisy Intermediate-Scale Quantum) architectures. Since the objective function is estimated from quantum measurements, finite sampling can also introduce statistical uncertainty into the classical optimization process. In addition, variational optimization landscapes may exhibit barren plateaus, in which parameter gradients become exponentially small with the number of qubits \cite{McClean2018,Scriva2024,Kungurtsev2024}.

In the present work, QAOA is included as a variational baseline for comparison with the feedback-based approaches. The main methodological distinction considered here is how the circuit parameters are determined: in QAOA, they are obtained through an external classical optimization procedure, whereas in FALQON the control parameter at each layer is determined directly from measured observables through a prescribed feedback law. Therefore, the comparison with QAOA provides a reference for contrasting variational and feedback-based optimization strategies, while the primary focus of this work remains on FALQON and its feedback-based variants.

The FALQON algorithm \cite{Magann_2022} was proposed to overcome
the dependence on an external classical parameter optimizer by using system measurements to dynamically update the circuit and parameters according to a feedback law. This distinction does not remove the measurement requirement itself: the expectation values used to determine the feedback parameters must still be estimated from quantum measurements and are therefore subject to finite-shot uncertainty \cite{Scriva2024,Barligea2025}.
This strategy is grounded in quantum Lyapunov control techniques, which can promote, under appropriate step-size conditions, a monotonic descent trajectory of the expected energy value across circuit layers.

\subsection{Formalism and Dynamics}

Considering the Ising Hamiltonian already formulated to encode the DNA assembly problem,

\[
H_p
=
\sum_{i<j}J_{ij}Z_iZ_j
+
\sum_i h_i Z_i,
\]
FALQON employs as a driver Hamiltonian the transverse field term

\begin{equation}
H_d
=
\sum_{k=1}^{n}X_k,
\label{eqn:Hd}
\end{equation}
where \(n\) is the number of qubits associated with the QUBO instance.

The system evolution is implemented through a sequence of alternating layers, applied according to the Trotter--Suzuki approximation:

\[
|\psi_{k+1}\rangle
=
e^{-i\beta_k H_d\Delta t}
e^{-iH_p\Delta t}
|\psi_k\rangle,
\]
where \(\Delta t\) is a fixed time step and \(\beta_k\) is the only parameter adjusted at each layer.

Unlike QAOA, the parameters \(\beta_k\) are not optimized by classical algorithms but dynamically updated through a feedback law based on the expectation value of the commutator between the Hamiltonians:

\begin{equation}
\beta_{k+1}
=
-i\langle\psi_k|[H_d,H_p]|\psi_k\rangle.
\label{eqn:beta1}
\end{equation}

This procedure implements a Lyapunov control where the feedback term enforces the time derivative of the expected energy \(\langle H_p\rangle\) to be non-positive, promoting a monotonic energy descent trajectory along the iterations \cite{Magann_2022}.

\subsection{Characteristics and Convergence}

The FALQON protocol guarantees that, under regular conditions on the step size \(\Delta t\), the expectation value of the energy \(J_k=\langle\psi_k|H_p|\psi_k\rangle\) decreases with each applied layer, driving the system to progressively lower-energy states.

 The feedback law removes the need for an external classical parameter-optimization routine, while the measurement requirements of the protocol remain relevant.

In the context of DNA reconstruction, the application of FALQON makes it possible to search for optimal assembly configurations without relying on classical parameter optimization, which motivates its investigation for depth-constrained quantum optimization. The present simulations, however, do not establish robustness to hardware noise or scalability to medium- or large-scale genomic instances.

\section{TR-FALQON Algorithm}
\label{sec:tr-falqon}

\subsection{Temporal Rescaling and TR-FALQON Formalism}

The \textit{temporal rescaling} method redefines the evolution of physical time $t$ in terms of a new parameter $\tau$ through the transformation $t=f(\tau)$, where $f(\tau)$ is a monotonically increasing function that satisfies $f(0)=0$ and $f(\tau_f)=t_f$, and it is referred to as the temporal rescaling function \cite{Rattighieri2025}. By rewriting the dynamics in terms of $\tau$,

$$
i\frac{d}{d\tau}|\psi(\tau)\rangle
=
H(f(\tau))\dot{f}(\tau)|\psi(\tau)\rangle,
$$
where $\dot{f}(\tau)=df(\tau)/d\tau$ and $\mathcal{H}=H(f(\tau))\dot{f}(\tau)$ represents the rescaled Hamiltonian. The term $\dot{f}(\tau)$ acts as a dynamic scaling factor: when $\dot{f}(\tau)>1$, the evolution is accelerated relative to physical time, while $\dot{f}(\tau)<1$ slows down the evolution, allowing greater resolution in critical regions of the dynamics. Thus, rescaling directly controls the ``rate'' at which each part of the evolution occurs.

To implement this dynamics in a layer-based algorithm, the evolution is discretized into steps of size $\Delta\tau$ and the Trotter--Suzuki decomposition \cite{Suzuki1990} is applied to approximate the evolution operator at each step. Each TR-FALQON layer takes the form

$$
|\psi_{k+1}\rangle
=
e^{-i\beta_k H_d\dot{f}(k\Delta\tau)\Delta t}
e^{-iH_p\dot{f}(k\Delta\tau)\Delta t}
|\psi_k\rangle,
$$
where the factor $\dot{f}(k\Delta\tau)$ modifies the operators relative to the original FALQON.

The adaptation of the control parameters is designed to preserve the monotonic energy-descent condition and is given by

\begin{equation}
\beta_k
=
-
\frac{1}{\dot{f}(k\Delta\tau)}
\langle\psi_{k-1}|i[H_d,H_p]|\psi_{k-1}\rangle.
\label{eqn:beta_tr}
\end{equation}

Typical functions for $f(\tau)$ include polynomial and trigonometric forms \cite{Rattighieri2025}.

\subsection{Impact and Results}

 Previous numerical studies of TR-FALQON report faster convergence and reductions in the number of feedback layers relative to standard FALQON for selected optimization problems \cite{Rattighieri2025}. This motivates its evaluation here in a DNA-assembly setting, while hardware-level circuit depth and noise resilience are not assessed in the present simulations.

\section{SO-FALQON Algorithm}
\label{sec:so-falqon}

\subsection{Mathematical Formalism}

In SO-FALQON, the control parameter $\beta_k$ is updated based on the condition that the expected value of the energy decreases with each new circuit layer,

\[
\Delta\langle H_p\rangle_k
=
\langle H_p\rangle_k
-
\langle H_p\rangle_{k-1}.
\]

To obtain this update, a second-order Taylor expansion of the evolution operators associated with $H_p$ and $H_d$ is performed. This expansion considers not only linear terms but also quadratic terms in $\beta_k$ and $\Delta t$, resulting in

\begin{equation}
\Delta\langle H_p\rangle_k
\simeq
\Delta t\,\beta_k A_{k-1}
+
(\Delta t)^2
\left[
\frac12\beta_k^2B_{k-1}
+
\beta_k C_{k-1}
\right],
\label{eqn:deltahp}
\end{equation}

where

\begin{align}
A_{k-1}
&=
i\langle\psi_{k-1}|[H_d,H_p]|\psi_{k-1}\rangle,
\nonumber\\
B_{k-1}
&=
\frac12
\langle\psi_{k-1}|[[H_d,H_p],H_d]|\psi_{k-1}\rangle,
\\
C_{k-1}
&=
\langle\psi_{k-1}|[[H_d,H_p],H_p]|\psi_{k-1}\rangle.
\nonumber
\end{align}

Minimizing expression~\eqref{eqn:deltahp} and initially assuming $B_{k-1}>0$, we obtain

\[
\beta_k
=
-\frac{2(A_{k-1}+\Delta t\,C_{k-1})}
{\Delta t\,B_{k-1}}.
\]

 Within the second-order approximation and under the stated assumptions, this choice is designed to satisfy $\Delta\langle H_p\rangle_k<0$.

In the case where $B_{k-1}<0$, in order to maintain the condition $\Delta\langle H_p\rangle_k<0$, the sign of $B_{k-1}$ is inverted, resulting in

\begin{equation}
\beta_k
=
-\frac{2(A_{k-1}+\Delta t\,C_{k-1})}
{\Delta t\,|B_{k-1}|}.
\label{eqn:beta2}
\end{equation}

In addition, it is common to impose bounds on the magnitude of $\beta_k$ to ensure numerical stability and maintain the validity of the Taylor expansion used \cite{Arai2025}.

The circuit is built with layers implemented through the Trotter--Suzuki decomposition \cite{Suzuki1990}, such that each step applies the operators $e^{-i\beta_kH_d\Delta t}$ and $e^{-iH_p\Delta t}$, incorporating the second-order feedback in each iteration.

\subsection{Practical Implementation and Results}

SO-FALQON has been demonstrated numerically in problems such as MAX-CUT and ground-state preparation, showing a reduction in the number of feedback layers required for convergence compared to first-order FALQON \cite{Arai2025}. Such layer reductions are potentially relevant to depth-constrained implementations, but they do not necessarily demonstrate reduced hardware noise or scalability to larger problem instances.

\section{Implementation}
\label{sec:implementacao}

The computational implementation of the proposed pipeline was carried out in Python, using a custom library developed in PyTorch \cite{paszke2019pytorch}, organized into modular stages, from DNA read preprocessing to the simulation and analysis of quantum optimization algorithms.

The starting point consists of reading and cleaning the DNA reads, followed by a systematic calculation of overlaps between each pair of reads, determining the largest suffix that coincides with the prefix of another read. In the experiments reported here, these overlaps are computed through exact suffix--prefix matches; an explicit sequencing-error model is not included in the overlap calculation as mentioned previously.

For the main SARS-CoV-2 benchmark, long reads were used to obtain more informative and less ambiguous overlaps: longer sequences reduce spurious matches and the degeneracy of the overlap scores, yielding a cleaner QUBO energy landscape. Because the encoding depends on the number of reads rather than on their length, this choice does not change the qubit footprint. The resulting benchmark remains tractable but does not reproduce the full error profile or ambiguity of large sequencing datasets.

This procedure is applied to all combinations of reads, constructing an overlap matrix (see Fig. \ref{fig:overlap_graph}) that will be used to define the weights in the QUBO Hamiltonian.

The overlap matrix is then used to construct the QUBO cost function, which is subsequently mapped onto an Ising-type Hamiltonian. This Hamiltonian is optimized using the FALQON, TR-FALQON, and SO-FALQON implementations described below.

For all simulations, the initial state was

\begin{equation}
|\psi_0\rangle
=
|+\rangle^{\otimes n_q},
\end{equation}
where $n_q$ denotes the number of qubits associated with the QUBO instance. The feedback-based methods used the transverse-field driver Hamiltonian defined in Eq.~\eqref{eqn:Hd}.

For comparison with the feedback-based approaches, QAOA was implemented as a variational baseline using the same problem Hamiltonian $H_p$, driver Hamiltonian $H_d$, and initial state employed in the FALQON simulations. The QAOA circuit was constructed with $p=300$ optimization layers, resulting in $2p$ variational parameters, $\boldsymbol{\gamma}$ and $\boldsymbol{\beta}$. The initial parameters were defined considering a linear schedule, $\gamma_k^{(0)}=0.13s_k$ and $\beta_k^{(0)}=0.13(1-s_k)$, with $s_k=(k+1/2)/p$. The parameters were optimized using the Adam optimizer \cite{Kingma2015Adam}. A short calibration stage of 10 optimization steps was performed for learning rates $0.015$, $0.03$, $0.05$, $0.07$, and $0.10$, from which $0.05$ produced the lowest energy and was therefore selected for the main optimization. The subsequent QAOA optimization was performed for 300 iterations. All QAOA simulations were carried out through exact state-vector evolution, without finite-shot sampling, and the energy gradients were evaluated using an adjoint-gradient procedure.

\subsection{Execution of Feedback-Based Algorithms}

Three algorithms were implemented: FALQON, TR-FALQON, and SO-FALQON. The main aspects of each variant are presented below.

\paragraph{FALQON (First Order).}
In standard FALQON, the update of the control parameter $\beta_k$ at each layer follows the first-order feedback rule described by Eq.~\eqref{eqn:beta1}.

\paragraph{TR-FALQON (Time-Rescaled FALQON).}

For TR-FALQON, the following temporal rescaling function was adopted,

\begin{equation}
f(\tau)
=
a\tau
-
\frac{t_f}{2\pi a}(a-1)
\sin\left(
\frac{2\pi a}{t_f}\tau
\right),
\label{eq:frt1}
\end{equation}
where $a$ is the temporal contraction parameter, $t_f$ is the final time, and $\tau$ represents the temporal variable associated with the layer. Owing to this choice, the temporal rescaling results in a contracted final time given by $\tau_f=t_f/a$. This function was chosen because it is the main one presented in the original work on the temporal rescaling method \cite{PhysRevResearch.2.013133} and also the primary choice in the article introducing the TR-FALQON algorithm \cite{Rattighieri2025}.

The rescaling factor $\dot{f}(\tau)$ is given by

\begin{equation}
\dot{f}(\tau)
=
a
-
(a-1)
\cos\left(
\frac{2\pi a}{t_f}\tau
\right).
\label{eq:dfrt1}
\end{equation}

At each layer $k$, the update of the control parameter is performed according to Eq.~\eqref{eqn:beta_tr}.

\paragraph{SO-FALQON (Second Order).}

In SO-FALQON, a hybrid approach was used for the update of $\beta_k$, computing both the first-order feedback rule $\beta^{(1)}$, described by Eq.~\eqref{eqn:beta1}, and the second-order feedback law $\beta^{(2)}$, described by Eq.~\eqref{eqn:beta2}.

After calculating both candidates for $\beta_k$, the algorithm selects between $\beta^{(1)}$ and $\beta^{(2)}$ according to the magnitude of each term,

\[
\beta
=
\begin{cases}
\beta^{(2)}, & \text{if }|\beta^{(1)}|>|\beta^{(2)}|,\\
\beta^{(1)}, & \text{otherwise}.
\end{cases}
\]

This adaptive choice allows SO-FALQON to combine the robustness of the second-order approximation with the stability of the first-order rule, resulting in efficient convergence even for relatively large time intervals, as detailed in \cite{Arai2025}.

\subsection{Metrics, Visualization, and Reproducibility}

At the end of the executions, the solutions are evaluated in terms of final energy, success probability of the optimal assembly, and convergence. All functions are parameterized and documented, enabling the replication of experiments and the extension to new datasets or encoding strategies.

\section{Results}
\label{sec:resultados}
\FloatBarrier

\begin{figure*}[!htbp]
    \centering
    \includegraphics[width=1\textwidth]{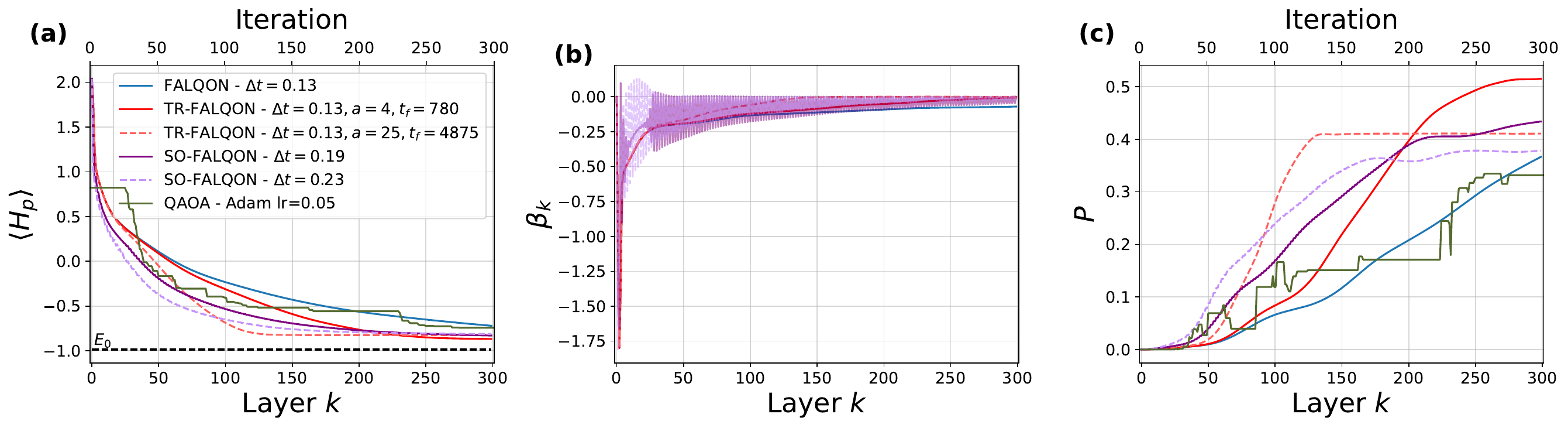}
    \caption{Comparison among feedback-based quantum algorithms and the QAOA variational baseline for the DNA assembly problem. Panel (a) shows the expected energy $\langle H_p\rangle$ obtained during the evolution of the feedback-based algorithms and during the QAOA optimization. Panel (b) shows the feedback control parameter $\beta_k$ for FALQON, TR-FALQON, and SO-FALQON; QAOA is not included in this panel because its variational parameters are independently optimized and are not directly equivalent to the FALQON feedback parameter. Panel (c) shows the probability of measuring the state encoding the optimal solution. For the feedback-based methods, the lower horizontal axis represents the circuit layer $k$. For QAOA, the upper horizontal axis in panels (a) and (c) represents the Adam optimization iteration. The QAOA curves report the best energy reached up to each optimization iteration and the success probability associated with the corresponding parameter set. The black dashed line in panel (a) indicates the ground-state energy $E_0$ of $H_p$.}
    \label{fig:falqon_comparacao}
\end{figure*}

The experiments were conducted to compare the performance of FALQON, TR-FALQON, and SO-FALQON in DNA assembly using a five-read SARS-CoV-2 instance. QAOA was additionally included as a variational baseline, allowing the feedback-based dynamics to be contrasted with an optimization strategy based on externally optimized circuit parameters. Under the one-hot encoding described in Sec.~\ref{sec:qubo}, this instance requires 16 qubits.

Figure~\ref{fig:falqon_comparacao} compares FALQON with two configurations of TR-FALQON and SO-FALQON. The configurations using the first hyperparameter set are denoted as TR-FALQON 1 and SO-FALQON 1 and are represented by the red and purple solid lines, respectively. The second configurations, TR-FALQON 2 and SO-FALQON 2, correspond to the dashed red and purple lines. The hyperparameter sets used for TR-FALQON and SO-FALQON are not identical: the first set employs milder parameters, closer to those used in FALQON, whereas the second adopts more aggressive values intended to promote faster convergence.

The QAOA baseline is represented by the green curve in panels (a) and (c). The feedback-based algorithms were evaluated for up to 300 circuit layers, whereas QAOA used a fixed depth of $p=300$ , where each QAOA layer consists of one evolution generated by the problem Hamiltonian $H_p$ and one evolution generated by the driver Hamiltonian $H_d$, and was optimized for up to 300 iterations. Unlike the feedback-based methods, for which each point corresponds to the state obtained after a given number of feedback layers, each QAOA point represents the best solution found up to a given optimization iteration. Therefore, the two horizontal axes should not be interpreted as equivalent measures of computational cost. Within these respective budgets, all methods reduced the expected energy, while their success-probability trajectories evolved differently.

Among the feedback-based configurations, TR-FALQON 1 reaches the lowest final energy and the highest success probability, exceeding $50\%$. TR-FALQON 2 and the two SO-FALQON configurations reach similar final energies but exhibit different success probabilities, while standard FALQON shows slightly weaker performance for this instance. TR-FALQON 2 reaches its minimum energy with the smallest number of layers among the tested configurations, and as its control parameter $\beta_k$ approaches zero, the success probability saturates at approximately $40\%$. Overall, the milder hyperparameter configurations produce a smoother initial convergence, closer to that of standard FALQON, whereas the more aggressive configurations lead to a steeper initial decrease followed by earlier stagnation.

The QAOA baseline also exhibits a progressive reduction of the objective energy during the classical optimization. Its final energy is lower than that of standard FALQON but remains above those obtained by the best-performing TR-FALQON and SO-FALQON configurations. Its success probability reaches approximately one third, remaining below the final values obtained by FALQON and by the tested TR-FALQON and SO-FALQON configurations. The step-like behavior of the QAOA curves reflects the fact that only improvements in the best energy found during the Adam optimization are displayed.

The comparison with QAOA further highlights the different optimization mechanisms involved. The feedback-based methods construct the control sequence sequentially from measured observables, whereas QAOA relies on the external optimization of a $2p$-dimensional parameter vector. Accordingly, the comparison in Fig.~\ref{fig:falqon_comparacao} is primarily intended to provide a variational reference for the obtained solution quality rather than a direct time-to-solution comparison. To complement this algorithmic comparison, we additionally estimate the two-qubit gate resources associated with the corresponding circuit constructions.

For this 16-qubit instance, the problem Hamiltonian contains 66 nonzero $ZZ$ interactions per layer, corresponding to 66 logical $R_{ZZ}$ operations. To obtain a hardware-oriented estimate, calibration data were retrieved through Qiskit from the IBM \texttt{ibm\_kingston} backend \cite{JavadiAbhari2024,IBMQuantumBackend}. Using a two-CZ equivalent for each logical $R_{ZZ}$ operation, each problem layer corresponds to 132 two-qubit gates. Consequently, the 300-layer circuit contains 19,800 logical $R_{ZZ}$ operations, or 39,600 CZ-equivalent gates. At the time of the analysis, the median calibrated CZ duration was $68$ ns, while the backend-wide median coherence times were $T_1=228.90~\mu\mathrm{s}$ and $T_2=147.26~\mu\mathrm{s}$. Under a fully serial two-qubit-gate assumption, the deepest 300-layer circuit therefore corresponds to an accumulated two-qubit gate time of approximately $2.69$ ms, equivalent to approximately $11.76\,T_1$ and $18.29\,T_2$. These ratios should not be interpreted as the actual scheduled circuit duration, since independent two-qubit operations can be executed in parallel and the present estimate does not include routing or scheduling effects.

The same counting model provides a simplified comparison of the cumulative two-qubit resource cost of FALQON and QAOA. FALQON constructs the feedback sequence through circuits of progressively increasing depth, from one to 300 layers. Considering one circuit evaluation at each depth, the accumulated sequence contains 2,979,900 logical $R_{ZZ}$ operations, corresponding to an estimated serial two-qubit gate time of approximately $0.405$ s. In contrast, a single QAOA circuit with 300 layers contains the same 19,800 logical $R_{ZZ}$ operations as the deepest FALQON circuit. Assuming one evaluation of this fixed-depth circuit per Adam iteration, the 300 optimization iterations correspond to 5,940,000 logical $R_{ZZ}$ operations and approximately $0.808$ s of serial two-qubit gate time. Thus, within this restricted metric, the cumulative two-qubit gate-time estimate for QAOA is approximately twice that of FALQON. This comparison is restricted to the two-qubit gate-counting model considered here and should not be interpreted as a direct hardware time-to-solution measurement.

In addition to the main SARS-CoV-2 instance, we carried out supplementary experiments with $n=\{4,5,6\}$ reads to probe how performance changes as the problem size increases. Across these finite-size instances, the maximum success probabilities decrease as the number of reads increases, with the most pronounced reduction observed for the six-read instance. Although this behavior is consistent with the increasing size of the encoded search space, the limited number of small instances considered here, together with the quadratic qubit overhead of the one-hot encoding, does not support an extrapolation to asymptotic scaling or biologically large instances. These additional benchmarks therefore provide a finite-size probe rather than a demonstration of scalability. Their full results are reported in Appendix~\ref{app:mito_results}.

\section{Conclusion}
\label{sec:conclusao}

In this work, we investigated the application of the FALQON algorithm and its variants TR-FALQON and SO-FALQON to the DNA assembly problem formulated in the QUBO/Ising formalism. The results show that, although the original FALQON provides a feedback-based optimization procedure without an external classical parameter-optimization routine and can promote monotonic energy reduction under appropriate step-size conditions, its circuit depth still imposes practical limitations on NISQ devices.

The considered variants demonstrated relevant advantages for the DNA-assembly instances considered in this work: TR-FALQON, by introducing temporal rescaling, achieved high success probabilities with fewer circuit layers than standard FALQON in selected configurations, while SO-FALQON maintained stable convergence for the tested larger time intervals and reduced the number of feedback layers required to reach high-quality solutions in selected configurations.

The inclusion of QAOA as a variational baseline further provides a reference for the solution quality obtained by the feedback-based approaches on the same DNA-assembly instance. Within the respective computational budgets considered, the best-performing TR-FALQON and SO-FALQON configurations reached lower final energies than the optimized QAOA circuit, while standard FALQON finished at a higher energy. All tested feedback-based methods reached higher final success probabilities than QAOA. However, the QAOA results are indexed by classical Adam optimization iterations for a fixed 300 circuit layers, whereas the feedback-based results are indexed by circuit layers. Therefore, the optimization trajectories should be interpreted primarily in terms of the obtained solution quality, while the two-qubit gate analysis reported in Sec.~\ref{sec:resultados} provides a simplified resource estimate rather than a direct hardware time-to-solution comparison.

These results indicate that enhanced feedback strategies, such as those incorporated in TR-FALQON and SO-FALQON, represent promising directions for feedback-based quantum computing in combinatorial problems.

In particular, the reduction in the number of layers observed in the studied instances suggests a potential advantage for depth-constrained quantum implementations. However, these results do not establish a quantum speedup over classical optimization methods, nor do they demonstrate scalability to biologically large genome-assembly instances. The present study should therefore be interpreted as a proof-of-concept evaluation of feedback-based quantum optimization for reduced DNA-assembly problems. Extending these methods to larger instances will require further investigation of quadratic encoding overhead, sampling costs, hardware noise, and comparisons with competitive classical and quantum optimization approaches. We leave for future work the investigation of multibasis encodings \cite{carmo2025warmpce, padin2026pauli} and space-efficient encodings \cite{glos2022space}, which, when combined with our approach, could in principle mitigate some of these scalability limitations.
\backmatter

\bmhead{Acknowledgements}

P.M.P. acknowledges support from Funda\c{c}{\~a}o de Amparo {\`a} Pesquisa do Estado de S\~ao Paulo (FAPESP) through Grants No.~2023/12110-7 and No.~2025/15490-0 (Thematic Project), and from CENPES/Petrobras (Grant No.~2025/00615-2). G.S.F. (2025/19585-6), L.A.M.R. (2025/25024-7), R.S.C. (2026/00378-3), A.D. (2024/19054-8), and G.E.L.P. (2025/22498-8) acknowledge support from FAPESP. F.F.F. acknowledges partial financial support from the National Institute of Science and Technology for Applied Quantum Computing through CNPq (Process No.~408884/2024-0) and from FAPESP through the Center for Research and Innovation on Smart and Quantum Materials (CRISQuaM, Process No.~2024/00998-6). The authors are grateful to Rodrigo Ara\'ujo Siqueira Barreiro and Gabriela Chiuffa Tunes of the AAMO group for helpful discussions.

\section*{Declarations}

\bmhead{Conflict of interest}

The authors have no conflicts to disclose.

\bmhead{Author contribution}

P.M.P. and L.A.M.R. performed the experiments, analysed the results, and wrote the manuscript. R.S.C., G.S.F., and G.E.L.P. contributed to the discussion of the results and reviewed the manuscript. A.D. wrote and reviewed the manuscript. E.G.D. provided the data and reviewed the manuscript. T.F.A. reviewed the manuscript. F.F.F. conceived and supervised the study, contributed to the interpretation of the results, and wrote and reviewed the manuscript. All authors read and approved the final manuscript.

\bmhead{Data/code availability}

The data underlying this study and the code used to generate the reported results are available from the corresponding author upon reasonable request.

\begin{appendices}

\appendix

\section{Additional Benchmarks: Human Mitochondrial Reads (4--6)}
\label{app:mito_results}
\FloatBarrier

The experiments in this appendix compare the performance of FALQON, TR-FALQON, and SO-FALQON in DNA assembly using sets of 4, 5, and 6 sequence fragments extracted from the human mitochondrial reference genome (NC\_012920.1) \cite{GenBankNC012920}; the sequences are listed in Table~\ref{tab:app:reads}. To satisfy the \textit{one-hot} constraint and ensure combinatorial feasibility, the QUBO Hamiltonian was constructed according to the formalism proposed by Lucas \cite{Lucas2014}. To reduce the number of required qubits, the read with index 0 in each set was fixed in the first position. In this way, the problem encoding requires $(n-1)^2$ qubits for $n$ reads, resulting in 9, 16, and 25 qubits for the sets of 4, 5, and 6 reads, respectively. For clarity, we denote the number of qubits by $n_q=(n-1)^2$ in the state-preparation expressions below.

\begin{table}[!ht]
\centering
\begin{tabular}{ccl}
\hline\hline
$\quad$\textbf{Reads}$\quad$
&
$\quad$\textbf{Index}$\quad$
&
\textbf{Sequence}$\quad$
\\
\hline
\multirow{4}{*}{4}
& 0 & {\texttt{ATGGCGTGCA}}\\
& 1 & {\texttt{GCGTGCAATG}}\\
& 2 & {\texttt{TGCAATGGCG}}\\
& 3 & {\texttt{AATGGCGTGC}}\\
\hline
\multirow{5}{*}{5}
& 0 & {\texttt{ATGACCAACAACCTC}}\\
& 1 & {\texttt{AACAACCTCGGGCCC}}\\
& 2 & {\texttt{CTCGGGCCCTGACGC}}\\
& 3 & {\texttt{CCCTGACGCCTACGC}}\\
& 4 & {\texttt{GCCTACGCTCCTGGC}}\\
\hline
\multirow{6}{*}{6}
& 0 & {\texttt{AGTGAAATTGACCTGCCCGTGAAGA}}\\
& 1 & {\texttt{CTGCCCGTGAAGAGGCGGGCATAAC}}\\
& 2 & {\texttt{CGGGCATAACACAGCAAGACGAGAA}}\\
& 3 & {\texttt{ACAGCAAGACGAGAAGACCCTATGG}}\\
& 4 & {\texttt{AAGACCCTATGGAGCTTTAATTTAT}}\\
& 5 & {\texttt{TTTAATTTATTAATGCAAACAGTAC}}\\
\hline\hline
\end{tabular}

\caption{ Sets of sequence fragments used in the additional experiments (human mitochondrial reference genome, NC\_012920.1) \cite{GenBankNC012920}.}
\label{tab:app:reads}
\end{table}

In all experiments, the initial state used was

\begin{equation}
\ket{\psi_0}
=
\ket{+}^{\otimes n_q},
\end{equation}

and the driver Hamiltonian adopted was

\begin{equation}
H_d
=
\sum_i X_i.
\end{equation}

\begin{figure*}[!htbp]
    \centering
    \includegraphics[width=\textwidth]{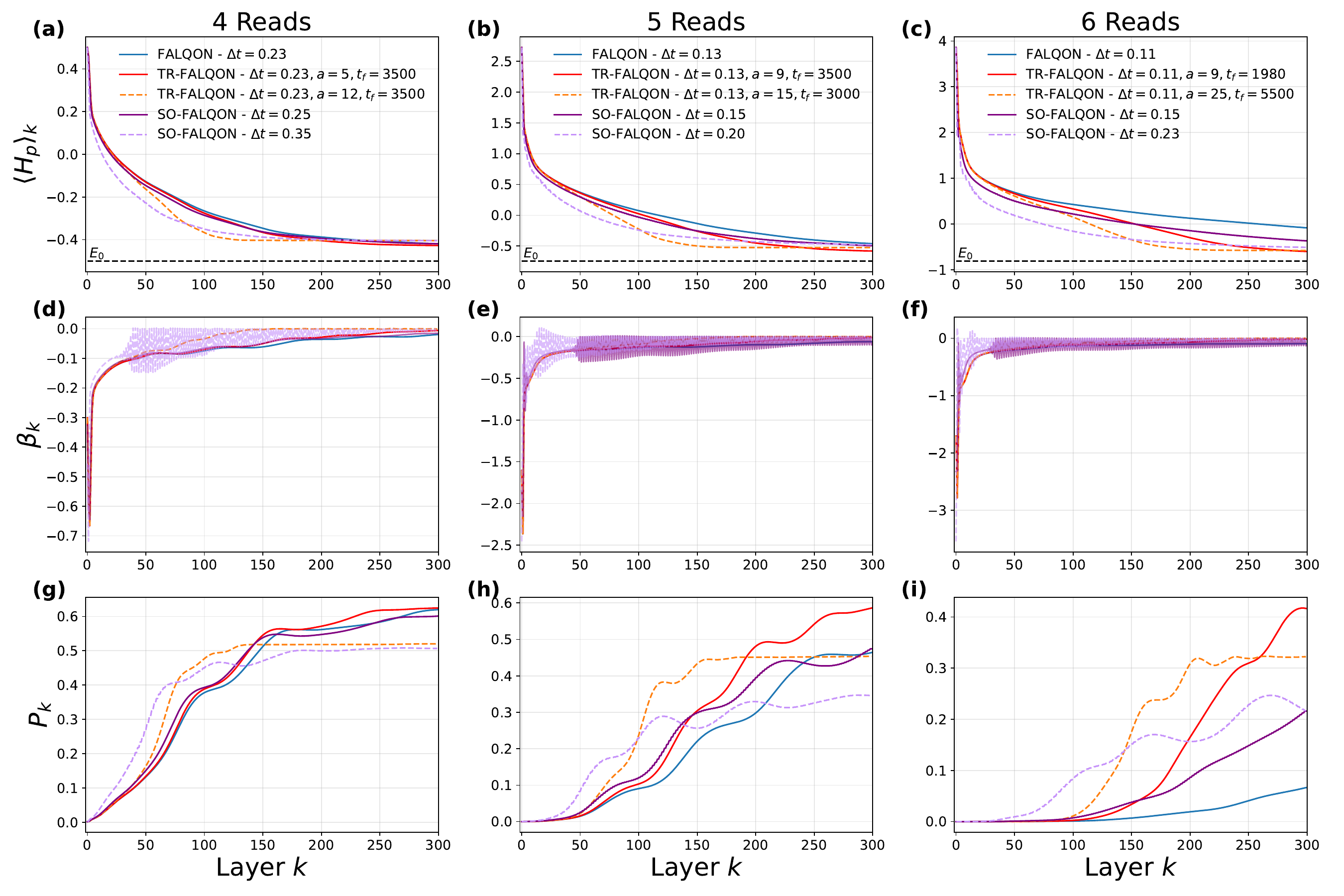}
    \caption{Performance of FALQON and its variants in the DNA assembly problem for instances with 4, 5, and 6 reads. The first row shows the expected value of the energy $\langle H_p\rangle_k$ as a function of the circuit layer $k$. The second row presents the values of the control function $\beta_k$. The third row displays the probability $P_k$ of measuring the state that encodes the optimal solution of the problem in each circuit layer. Each curve corresponds to a different variant of the algorithm, as indicated in the legend. The black dashed curve in the panels of the first row indicates the ground-state energy $E_0$ of $H_p$.}
    \label{fig:app:falqon_comparacao}
\end{figure*}

The numerical results are presented in Fig.~\ref{fig:app:falqon_comparacao}, organized into a grid of three rows by three columns (columns: 4, 5, and 6 reads). The first row shows the expected value $\langle H_p\rangle_k$ versus the number of layers, the second row shows the control parameters $\beta_k$ across layers, and the third row shows the success probability

\begin{equation}
P_k
=
\big|\langle\text{solution}|\psi_k\rangle\big|^2,
\end{equation}

computed from the state produced at each layer. Each panel compares FALQON with TR-FALQON and SO-FALQON (for two hyperparameter choices in the latter methods). All algorithms were executed with up to 300 layers.

In the case of FALQON, the curves correspond to the time step that produced the largest energy decrease within the 300-layer limit. For 4 reads, this time step was below the critical step $\Delta t_c$ (largest value that still guarantees monotonic convergence of the energy), since near $\Delta t_c$ the algorithm exhibited stagnation in energy reduction. For 5 and 6 reads, no stagnation was observed, enabling the direct use of $\Delta t_c$.

 Overall, the supplementary finite-size benchmarks show that TR-FALQON and SO-FALQON can improve convergence and/or success probability relative to standard FALQON for selected configurations, consistent with previous numerical studies of feedback-based variants \cite{Arai2025,Rattighieri2025,Magann_2022}. These results concern idealized simulations and should not be interpreted as evidence of noise resilience or scalability on current hardware.

\section{TR-FALQON Hyperparameter Sensitivity and Stagnation Analysis}

To evaluate the sensitivity of TR-FALQON to the time-rescaling parameters, we performed additional simulations using a five-read SARS-CoV-2 instance.

\begin{figure}[htp]
    \centering
    \includegraphics[width=1\linewidth]{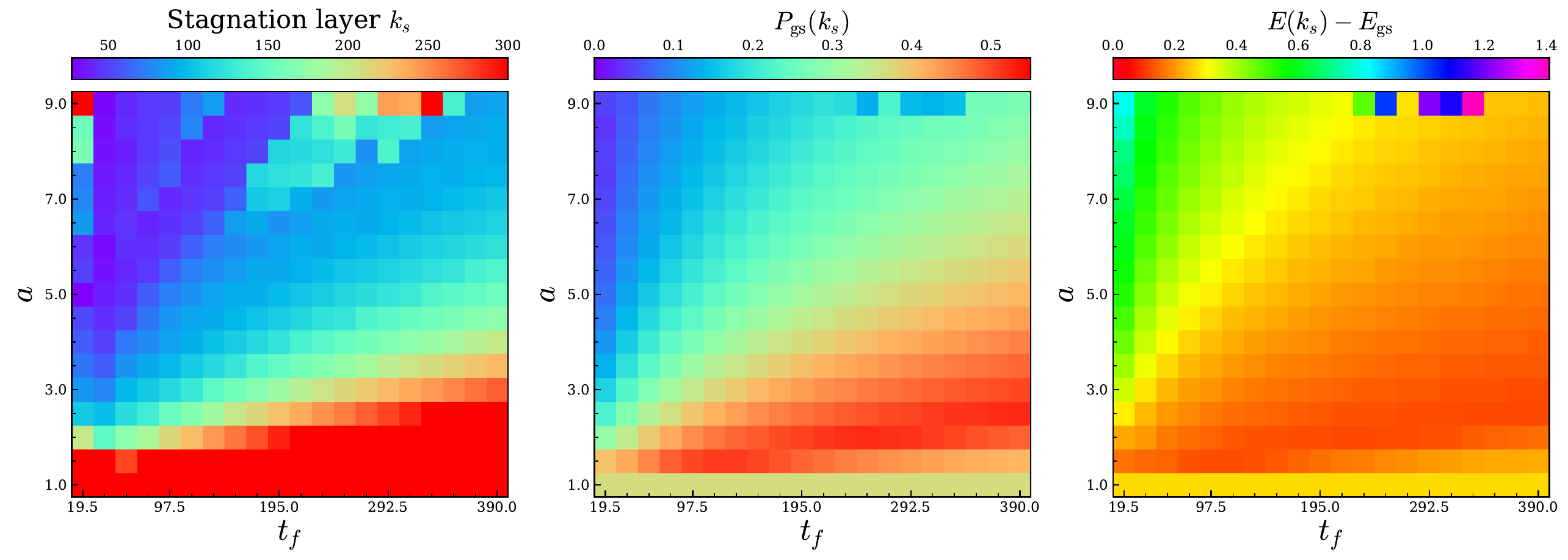}
    \caption{TR-FALQON hyperparameter analysis for the five-read SARS-CoV-2 instance with $\Delta t=0.13$. The heatmaps show the stagnation layer $k_s$, the success probability $P_{\mathrm{gs}}(k_s)$, and the energy difference $E(k_s)-E_{\mathrm{gs}}$ at the stagnation point for different values of $a$ and $t_f$.}
    \label{fig:tr-parameters}
\end{figure}

Figure \ref{fig:tr-parameters} shows the results obtained by varying $a$ from 1 to 9 and $t_f$ from $0.5\Delta t N_{\mathrm{layers}}$ to $10\Delta t N_{\mathrm{layers}}$. The time step was fixed at $\Delta t=0.13$, the same value used in the main analysis. For each combination of $a$ and $t_f$, we determined the stagnation layer $k_s$ and evaluated the success probability and energy reached at this point.

For $a=1$, the time-rescaling function reduces to the standard FALQON, independently of $t_f$. In this case, no stagnation is observed within the analyzed interval, but the success probability remains around $0.35$. For $a=1.5$, stagnation is also not observed, while the success probability increases to values around $0.4$. For $a=2$, the behavior becomes more dependent on $t_f$. Larger values of $t_f$ delay stagnation and lead to higher success probabilities than those obtained with standard FALQON.

When $a$ increases to approximately $2.5\lesssim a\lesssim4$, stagnation occurs earlier for several configurations. However, this does not necessarily correspond to worse results. In this region, the success probability increases and, near $a\approx3$, particularly for larger values of $t_f$, some of the highest probabilities in the map are obtained, approaching $0.5$. Thus, some configurations reach better states even when the evolution stagnates earlier. This can be seen in the region where stagnation occurs around 150 layers, while the success probability remains higher than that obtained with standard FALQON.

For $a\gtrsim5$, the behavior starts to change. In general, increasing $a$ increases the tendency toward stagnation, requiring larger values of $t_f$ to delay it. However, for large values of $a$, the time rescaling becomes strong enough to affect the convergence of the algorithm. This behavior is more evident near $a\approx9$, where the evolution loses its expected monotonic behavior and oscillations appear for some values of $t_f$. In this regime, the behavior is therefore not limited to earlier stagnation, since convergence itself can be lost. The success probability also decreases, remaining approximately between $0.2$ and $0.3$ for a large fraction of these configurations.

Considering the three maps together, the best results are mainly found for small and intermediate values of $a$ combined with larger values of $t_f$. These regions present both higher success probabilities and lower energies. However, this does not imply that $a=1$ is the best choice, since standard FALQON gives worse results than several configurations with $a>1$. Therefore, there is an intermediate time-rescaling regime in which TR-FALQON reaches better states, even when the evolution stagnates earlier.

We also observe a relation between energy and success probability. Regions with lower energies tend to coincide with regions of higher $P_{\mathrm{gs}}$. This relation is useful for hyperparameter selection because the success probability requires knowledge of the optimal solution, which is generally not available for an unknown problem. Although $E-E_{\mathrm{gs}}$ is shown in Fig. \ref{fig:tr-parameters} to facilitate comparison with the exact solution for this instance, the hyperparameters can be selected directly from the measured energy $E=\langle H_p\rangle$.

\end{appendices}

\newpage
\bibliography{bibl}

\end{document}